\documentclass[conference]{IEEEtran}
\usepackage{cite}
\usepackage{amsmath,amssymb,amsfonts,amsthm}
\usepackage{stmaryrd}
\usepackage{algorithmic}
\usepackage{graphicx}
\usepackage{textcomp}
\usepackage{xcolor}
\usepackage{makecell}
\usepackage{optidef}
\usepackage{algorithm}
\usepackage{algorithmic}
\usepackage{adjustbox}
\usepackage{flushend}

\usepackage{subcaption}
\usepackage{tabulary}

\def\BibTeX{{\rm B\kern-.05em{\sc i\kern-.025em b}\kern-.08em
    T\kern-.1667em\lower.7ex\hbox{E}\kern-.125emX}}

\newcommand{\norm}[2]{\ensuremath{\left\lVert #1 \right\rVert}_#2}

\SetSymbolFont{stmry}{bold}{U}{stmry}{m}{n}

\begin{document}

\title{Influence of Dataset Parameters on the Performance of Direct UE Positioning via Deep~Learning}

\author{
    \IEEEauthorblockN{Baptiste Chatelier, Vincent Corlay, Cristina Ciochina, Fallou Coly, and Julien Guillet}
    \IEEEauthorblockA{Mitsubishi Electric R\&D Centre Europe, Rennes, France. E-mail: v.corlay@fr.merce.mee.com}
}

\maketitle


\begin{abstract}
 User equipment (UE) positioning accuracy is of paramount importance in current and future communications standard. However, traditional methods tend to perform poorly in non line of sight (NLoS) scenarios. 
As a result, deep learning is a candidate to enhance the UE positioning accuracy in NLoS environments. 
In this paper, we study the efficiency of deep learning on the 3GPP indoor factory (InF) statistical channel. 
More specifically, we analyse the impacts of several key elements on the positioning accuracy: the type of radio data used,  the number of base stations (BS), the size of the training dataset, and the generalization ability of a trained model. 
\end{abstract}


\section{Introduction}
Recent advances in modern factories, such as the  industrial internet of things, have allowed the emergence of automated guided vehicles (AGV). 
Such AGV require extremely precise positioning in real-time.
However, indoor scenarios, and particularly indoor factories, are challenging for positioning: They are mainly NLoS scenarios. 

Unfortunately, conventional positioning techniques used in the 5G NR releases \cite{Pos_Dwivedi} perform poorly in NLoS scenarios: Typically, the 90$\%$ quantile of the positioning error\footnote{Called the $Q(0.9)$ value in this paper, see Section~\ref{Sec_sim_results} for the definition.} is greater than 15m, see Proposal~5.9-1 in \cite{TdocSumOct}.  
To address this issue, the radio acess nework working group 1 (RAN1) from 3GPP, the standardisation body for 5G, 
opened a study item on the use of machine learning techniques for UE positioning in InF scenarios \cite{SI_item}. A recent ``summary of evaluation on AI/ML for positioning accuracy enhancement" is available in \cite{TdocSumOct}. One can also refer to \cite{Pos_survey} for a survey on the use of AI/ML for Indoor Positioning. 

In this paper, we investigate the deep learning approach for direct UE positioning in highly NLoS scenarios. 
On the one hand, the performance of a deep learning model highly depends on the training dataset. 
On the other hand, there may be strong constraints on the data available for the training in a 5G context:  
sparse measurements in a given scene, specific or noisy radio signals used, etc. 
As a result, it is critical to assess the sensitivity and robustness of this positioning method to the dataset parameters.

Similarly to the 3GPP RAN1, data from statistical channels are considered to assess the deep learning efficiency, namely the 3GPP InF channel\cite{TR38901}. This framework is presented in Section~\ref{Section_intro}.
In Section~\ref{Sec_sim_results}, we then report simulation results to highlight the influence on the positioning performance of the four following aspects:
\begin{itemize}
\item The type of radio signal: We consider both the path gain (PG) and the channel impulse response (CIR) as input data to infer the associated position.
\item  The number of BS: The baseline performance are presented with 18 BS as described in the 3GPP statistical channel. We also investigate the positioning performance of the PG and CIR with a reduced number of BS.
\item The dataset size: We analyse the influence of the dataset size on the positioning accuracy. 
\item The generalization abilities of the trained models: We show the positioning performance of a model, trained with data from a specific factory, for a UE in another factory. We consider two cases: without fine tuning on the new factory and with fine tuning (in other words, transfer learning).
\end{itemize}

Some of these aspects have also been investigated in the scope of the 3GPP study on this topic, such as the positioning performance based on the CIR and a dataset of size of 80 000, as well as some aspects of the generalization abilities \cite{TR38901}. However, to the best of the author's knowledge, the performance with the PG, the influence of the number of BS, the influence of the dataset size, and the transfer learning abilities of the models have not yet been investigated. 

Note that some other works also consider alternative signals for UE localization in the InF scenario, such as \cite{InPos_bjorn} where the authors consider the carrier phase.

\begin{table*}
\vspace{2mm}
\begin{center}
\begin{tabular}{|c | c | c | c | c|} 
\hline
 & Number of data available & Number of scenes available &Model Accuracy & Model specificity  \\
\hline
True measurements & - & - - & +++ & - - - \\
\hline
Digital twin & +& + & ++ & -  \\
\hline
Statistical channel&  ++& ++  & + & + \\
\hline
\end{tabular}
\caption{Assessment of the possible approaches to generate datasets in order to train a neural network. The sign $+$ stands for strength and $-$ for weakness.}
\label{table_pro_con}
\end{center}
\end{table*}

\section{Presentation of the framework}
\label{Section_intro}

\subsection{Direct UE positioning}

We consider an uplink communication where a UE broadcasts a signal to several base stations (BS). 
The devices have only one antenna. 
The horizontal coordinates of a UE at position $k$ is denoted by  $\mathbf{p}_k=\left(x_k,y_k\right)$.

The goal is to train a neural network at the network side to output an estimate $\hat{\mathbf{p}}_k$ given the channel measurements made by the BS as input.
This approach is commonly called direct positioning or fingerprint positioning.

\subsection{Possible approaches to model a radio channel}

In order to train and assess the performance of a neural network, one needs to create a dataset of channel measurements.
Several approaches can be envisioned for this dataset creation. 

Firstly, measurements from true channels can be realized. These measurements can be performed offline by engineers or online by BS with the use of 5G NR positioning reference signals, collected e.g., via the location management function (LMF) of the 5G core network. 

A second approach consists in collecting measurements from a simulated radio channel in a digital twin. This requires implementing a digital twin of the considered scenario. The digital twin can be simplified or advanced. 

Finally, statistical channels can be used to generate the datasets. In this case, the channel parameters are random variables whose statistics are chosen to represent one category of channels. For instance, the 3GPP InF channel specifies specific probabilities, e.g., the LoS/NLoS status, which are consistent with the probabilities encountered in true factories. Then, one realization of this statistical channel represents one virtual factory. This approach is used in the study item of 3GPP mentioned in the introduction.

In Table~\ref{table_pro_con}, we rate the different approaches according to the following criteria: the number of data available, the number of different scenes available, the model accuracy, and the model specificity (whether the results apply only to the considered scenario or if they allow to establish rules valid for other scenarios). The choice of the model depends on the goal of the study: studying the neural network in a specific scenario, or in a variety of scenarios, pre-training a model, etc.

As an example, in~\cite{Arnold2018} pre-training of a model is done on a very simplified digital-twin (where only the path loss as a function of a distance  is implemented). Then, fine tuning is performed with real measurements. The drawback of this approach is the following: It is difficult to state if the results hold for any NLoS environment or just for the one used to collect the measurements

In this study, we use the statistical approach. As reported in Table~\ref{table_pro_con}, the main strength of this approach is to provide a large number of scenes from one category, with flexibility on the number of data per scene. It is thus suited to a study of the influence of several dataset parameters on the deep learning-based positioning accuracy. 
More specifically, the 3GPP InF scenario is considered. This latter scenario is selected because of its highly NLoS characteristics, see Figure~\ref{fig:LOS_NLOS_status} and related discussion in Section~\ref{sec_inf_spe}.

\subsection{Considered statistical channel}

We consider the 3GPP InF-DH scenario as described in TR 38.901 \cite{TR38901}. 
This is a highly NLoS scenario due to many (virtual) clutters. 

\subsubsection{3GPP channel model}

Similarly to standard approaches, the 3GPP statistical channel is modelled by three main components:
1-The path loss, which represents the attenuation due to the distance between the BS and UE. 2-The shadowing (large-scale) fading, caused by long-term fluctuations of the received power due to partial obstructions. 3-The fast (small-scale) fading due to multipath propagation.
Accordingly, the coefficients of the CIR are generated by three categories of variables:
\begin{itemize}
\item The LoS/NLoS propagation condition, modelled by a Bernoulli random variable. Given a LoS or NLoS realization for a position, the path loss formula is then a deterministic function of the distance and the frequency.
\item The large-scale parameters (LSP): They represent realizations of both the shadow fading and parameters used to generate the small-scale parameters, e.g., the delay spread, the angular spread, Ricean K factor. They can be understood as general statistical properties of the channel. 
\item The small-scale parameters (SSP): They are variables used to generate the clusters and rays, e.g., the delays, the cluster power, the arrival and departure angles. They represent detailed channel characteristics.
\end{itemize}
The clusters and rays (simulating a multipath propagation) are then used to generate the CIR channel coefficients.

On the one hand, the LSP and SSP are slowly varying parameters (both in time and space), e.g., order of the meter. 
On the other hand, theses parameters induce fast fading, e.g., order of the wavelength. 
This is interesting in the scope of the considered problem: 
If the resolution order of the positioning system is higher than the wavelength, the fast fading induced by the SSP parameters appears to be completely random. 
A positioning system based on the channel coefficients (i.e, the CIR) should therefore learn to ignore (or average) some of these variations.

An alternative would be to infer the position based on the LSP or SSP, which (again) are more slowly varying. 
However, since the SSP are detailed channel characteristics, it is difficult to deduce them from the channel coefficients. 
The LSP are more general characteristics which should be easier to obtain by averaging the channel coefficients over several dimensions.

As an example, let us define the the path gain (PG) as the shadowing minus the path loss (in dB). While this quantity has no physical reality, it can be understood as the average over the small-scale fading of the received power (assuming ideal omni directional antennas and unit transmit power). As this quantity is slowly varying, it should not appear as random to the system. Nevertheless, the PG does not provide an information as rich as the CIR, which may be required for very-accurate positioning. Consequently, it is interesting to compare positioning performance based on the PG and the CIR.

Spatial consistency, as specified in TR 38.901 \cite{TR38901}, is required for a positioning study. This means, for instance, that the autocorrelation of the shadow fading random variable should be an exponential random variable (see Eq. 7.4-5 in TR 38.901). In our implementation, the spatial consistency is enabled for the three categories of variables listed above. 
The model should also respect spatial coherence meaning that a random variable realization at a given position at two different instants should be the same or at least statistically dependent.



\subsubsection{InF specificity}
\label{sec_inf_spe}
\begin{figure}[t]
    \centering
    \includegraphics[scale=2]{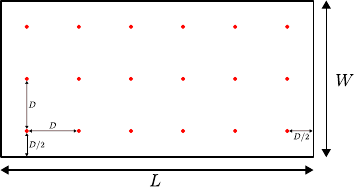}
    \caption{3GPP InF topology. BS are represented with red circles.}
    \label{fig:InF_topology}
\end{figure}

The topology of the considered factory  in the 3GPP InF model is a rectangular area of width $W=60$m, length $L=120$m and height $H=10$m. There are 18 BS at specific locations with inter-BS spacing $D=20$m. The BS height is $8$m. The UE height is fixed at $1.5$m. Figure~\ref{fig:InF_topology} shows this topology. 

In order to simulate various (virtual) objects in a factory that could interfere with the wireless propagation, the 3GPP InF model introduces clutter parameters $\left(r,h,d\right)$. 
These meta-parameters represent respectively the clutter density, height, and size. 
In practice, they influence the NLoS probability.
In this paper, two different clutter configurations are considered: $\left(r,h,d\right) = \left(40\%,2m,2m\right)$ and $\left(r,h,d\right) = \left(60\%,6m,2m\right)$. 

The radio parameters are set as follows: carrier frequency $f_c = 3.5$ GHz, bandwidth $W=100$ MHz. The BS and the UE antennas are omnidirectional with unitary gains.  

\begin{figure}
\centering
\begin{subfigure}{.5\columnwidth}
  \centering
  \includegraphics[width=.9\columnwidth]{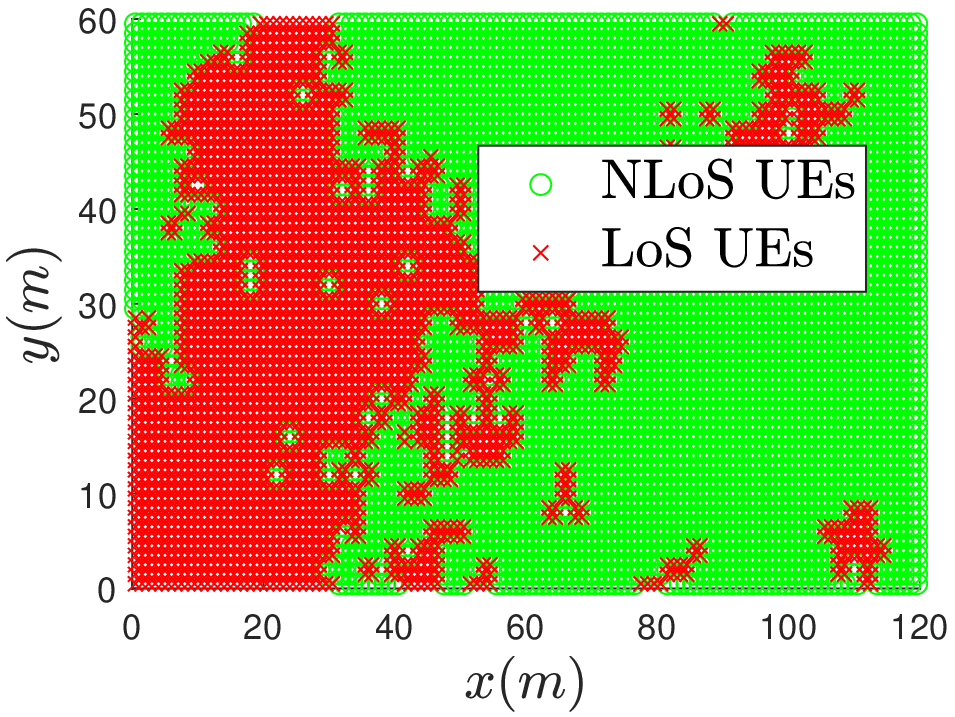}
  \caption{40$\%$}
\end{subfigure}%
\begin{subfigure}{.5\columnwidth}
  \centering
  \includegraphics[width=.9\columnwidth]{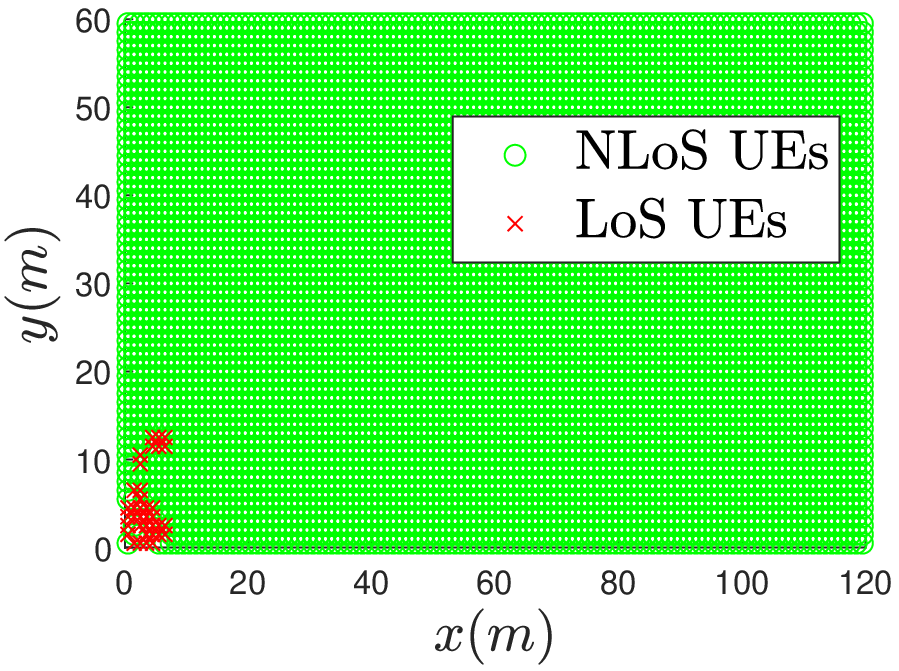}
  \caption{60$\%$}
\end{subfigure}
\caption{A LoS/NLoS realization for the 40$\%$ and 60$\%$ clutter density scenarios, respectively.}
\label{fig:LOS_NLOS_status}
\end{figure}

\begin{figure}[!h]
    \centering
    \includegraphics[scale=0.6]{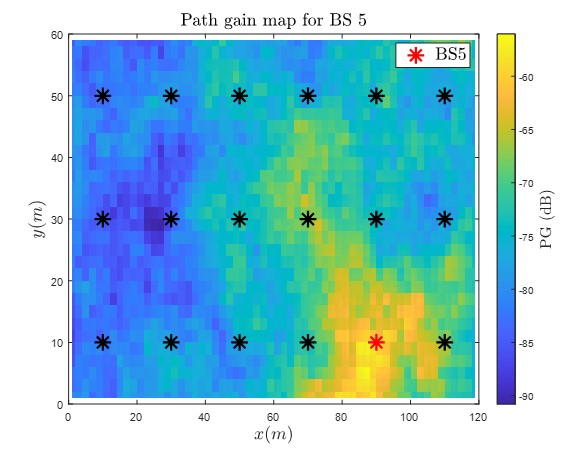}
    \caption{PG map from the perspective of the red BS.}
    \label{fig:PG_illu}
\end{figure}

As an example, Figure~\ref{fig:LOS_NLOS_status} shows a LoS/NLoS realization both for the 40$\%$ and 60$\%$ cases. 
Figure~\ref{fig:PG_illu} illustrates one realization of the PG from the perspective of one BS.

\subsection{Training and testing dataset generation}
\label{dataset}

Two categories of datasets are generated from from our system level simulator implementing the 3GPP InF-DH channel model and topology: PG datasets and  CIR datasets. 

For the PG datasets, for each position $k$, the PG at each of the 18 BS, $\mathbf{a}_k \in \mathbb{R}^{1\times 18} $, is computed and the horizontal coordinate of the position $\mathbf{p}_k \in \mathbb{R}^{1 \times 2} $ is associated as the label. 

The CIR datasets are generated in a similar manner, but where the data for each position is now $\mathbf{b}_k \in \mathbb{C}^{256\times 18}$ instead of $\mathbf{a}_k$, where 256 is the length of the CIR. 

If a dataset with a reduced number of BS is required, we simply down-sample these datasets and keep a reduced number of columns per sample, where a column of a sample represent one BS.

Regarding the positions in the dataset, the scene is uniformly sampled. For instance, to have a dataset of size 28~800, the inter-position spacing in both axes is 0.5m.

Once trained, the neural network is tested on positions not used for the training. 
In Sections~\ref{sec_type_radio_sig} and \ref{Section_dataset_size} signals from randomly sampled positions of the same factory are used: It means that the spatial consistency with the data in the training dataset is respected. 
In Section~\ref{sec_gene_sim}, where we study the generalization, signals from another factory are used: The spatial consistency with the training data is not respected.

\subsection{Neural network architectures}
\label{AI_arch}

\begin{figure}[t]
    \centering
    \includegraphics[scale=.8]{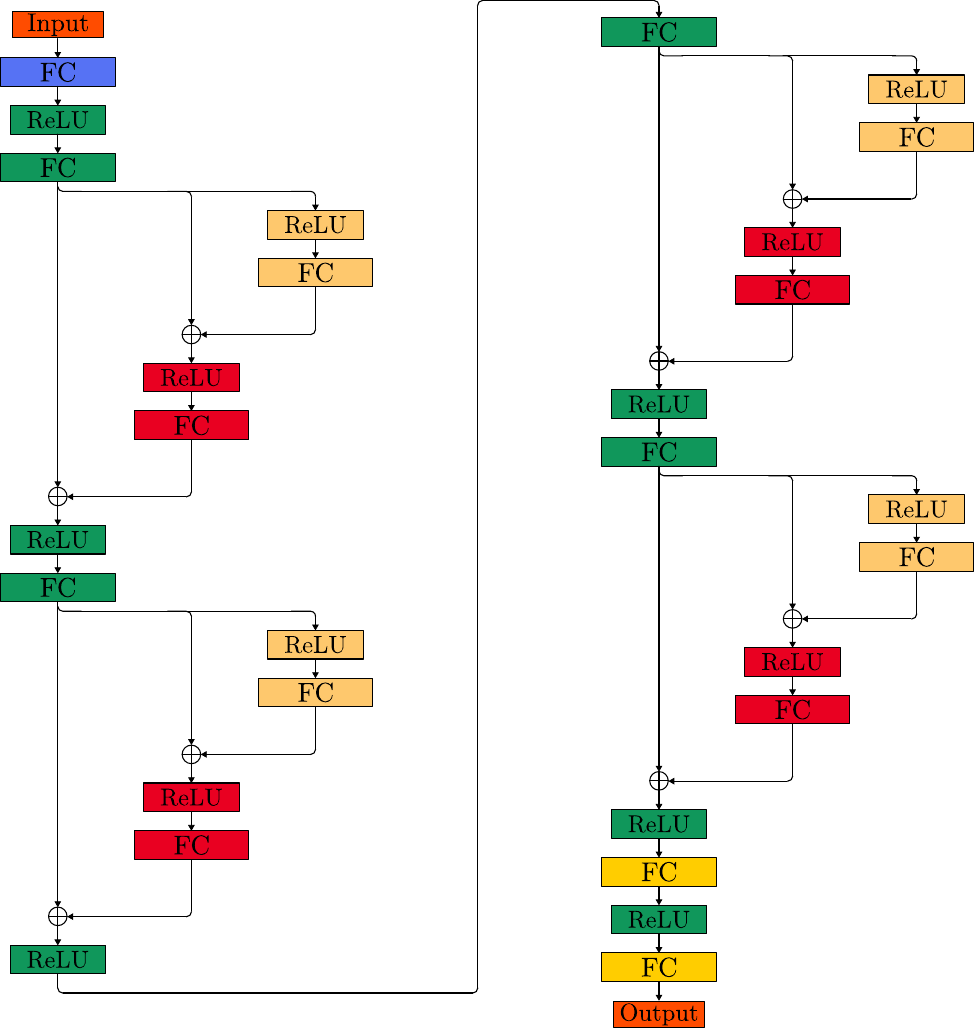}
    \caption{ResNet architecture for the PG datasets.}
    \label{fig:resnet_pg}
\end{figure}

The PG and CIR datasets are evaluated on neural networks with different architectures.
For the PG datasets, a customized residual network (ResNet) architecture is used, presented in Figure~\ref{fig:resnet_pg}. 
Each fully connected layer is composed of $120$ neurons. The neural network has 104.1k parameters.
For the CIR datasets, the ResNet18 architecture \cite{He2016} (with untrained weights) is considered. It has 11.1M parameters.

For the PG datasets, we choose a ResNet rather than a standard feed-forward neural network. Indeed, empirical results show that they perform better with the same number of parameters. 
For the CIR dataset, as the input has the structure of an image, a convolutional neural network (CNN) is adapted. 
The ResNet18 architecture, a CNN with a residual structure, offers good positioning performance for a relatively low number of parameters in comparison to other CNN such as ResNet50.

For both neural networks, the output is an estimated horizontal coordinate vector $\hat{\mathbf{p}}_k$. 
The loss function for the training is the mean squared error $\mathcal{L} = \frac{1}{2} \norm{\hat{\mathbf{p}}_k - \mathbf{p}_k}{2}^2$.

\section{Simulation results}\label{Sec_sim_results}



The main evaluation metric is the $90\%$ quantile of the cumulative distributive function (CDF) of the horizontal positioning error $Q\left(0.9\right)$, i.e., the value $Q\left(0.9\right)$ (in meters) such that $90\%$ of the errors are under $Q\left(0.9\right)$.

\subsection{Type of radio signal}
\label{sec_type_radio_sig}

We first compare the positioning performance using the PG and the CIR as input of the neural networks.
We use the baseline InF channel with 18 BS.
For both radio signals, the neural networks are trained with a dataset of 28 800 positions uniformly placed in the scene.

\begin{figure}[t]
    \centering
    \includegraphics[scale=.8]{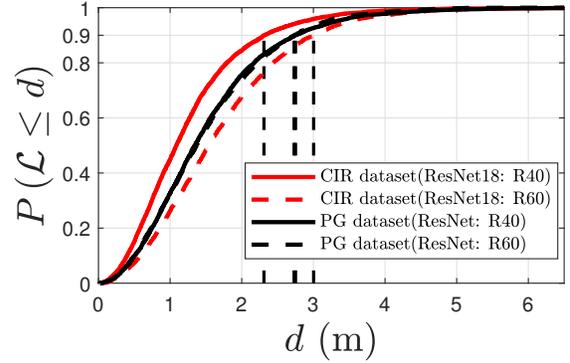}
    \caption{Training results for the PG and CIR datasets. The dataset size is 28 800. The $Q(0.9)$ values are: 2.4, 2.6, 2.6, and 3.0. }
    \label{fig:train_results_CIR_28_8k} 
\end{figure}

The results are provided in Figure~\ref{fig:train_results_CIR_28_8k}. 
There is no clear difference on the positioning accuracy when the inference is based on the PG or CIR. 
The observed differences may depend on more or less efficient hyper-parameter optimization. 
Indeed, for some dataset, the hyper-parameter optimization enabled to enhance the accuracy by 2m. 
Similarly, changing the density of clutters does impact significantly the results.

\subsection{Influence of the number of BS}

We also investigate the positioning performance based on the PG and CIR with a reduced number of BS.
As the CIR is richer (256 values per position and BS) than the PG (only 1 value per position and BS) we expect the former radio signal to be more robust to a reduced number of BS.

\begin{figure}
    \centering
    \includegraphics[scale=.95]{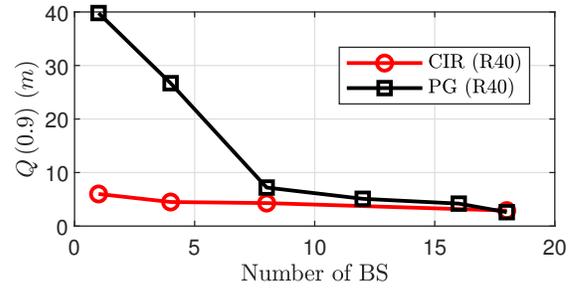}
    \caption{Influence of the number of BS on the positioning performance. }
    \label{fig:number_BS}
\end{figure}

The performance is provided on Figure~\ref{fig:number_BS}. This is consistent with what is expected. The positioning based on the PG requires at least 8 BS to get reasonable performance while only 1 BS is required for the CIR. Increasing the number of BS beyond these values only brings a marginal improvement.

\subsection{Influence of dataset size}
\label{Section_dataset_size}

\begin{figure}
    \centering
    \includegraphics[scale=.68]{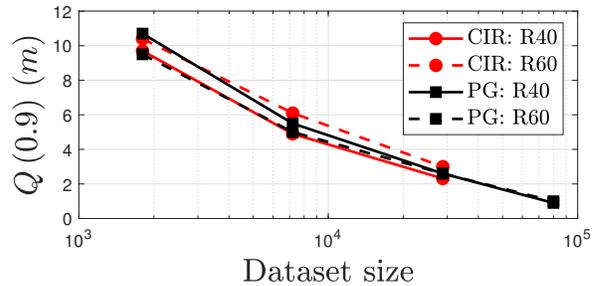}
    \caption{Influence of training dataset size on UE positioning performance. }
    \label{fig:dataset_comp}
\end{figure}

To study the influence of the dataset size on the UE positioning accuracy,
we generate training datasets of sizes going from 1700 to 80 000  for the PG and 1700 to 28 000 for the CIR, by changing the inter-position spacing. 
Results are presented in Figure~\ref{fig:dataset_comp}. 
As already observed with Figure~\ref{fig:train_results_CIR_28_8k}, there is not clear performance difference using the CIR or the PG with 18 BS.
Then, unsurprisingly, increasing the number of positions allows to obtain better UE positioning performance.
The $Q\left(0.9\right)$ value goes down from around 10m with a dataset of size 1700 to less than 1m with a dataset of  size 80 000. 
This latter result is consistent with the one reported in \cite{TdocSumOct}, where companies report $Q(0.9)$ values slightly below 1m or slightly higher for CIR-based positioning with a training dataset of size 80 000. 
Again, small differences compared to \cite{TdocSumOct} could be attributed to more or less efficient fine tuning of the hyper parameters but do not impact the following main observation on the dataset characteristics.

In the considered range, we have the following rule of thumb: The error is divided by 2 each time the size of the training dataset is multiplied by 4.
This is not surprising as reducing the 2D inter-position spacing by a factor 2 induces multiplying the dataset size by a factor 4.


\subsection{Generalization}
\label{sec_gene_sim}


We analyse the generalization performance, i.e., how a network trained on a specific factory (i.e., a channel realization) performs on another one.
As explained before, we consider two cases: without and with fine tuning.  

\begin{figure}[t]
    \centering
    \includegraphics[scale=.6]{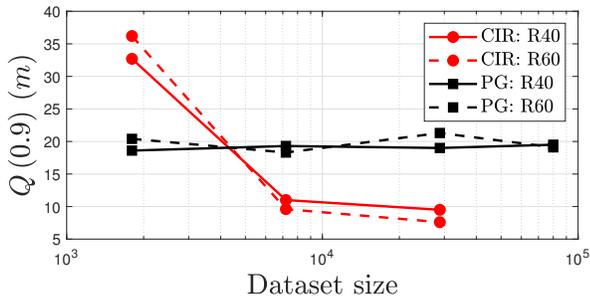}
    \caption{Influence of the training dataset size on UE generalization positioning performance.}
    \label{fig:dataset_comp_gene}
\end{figure}

The generalization performance without fine tuning and as a function of the training dataset size is shown on Figure~\ref{fig:dataset_comp_gene}. 
In this experiment, the neural networks are trained on a specific factory with a dataset of a given size and then tested on another factory.
Obviously, the CIR offers much better performance compared to the PG: The $Q(0.9)$ value is divided by two when the dataset is large enough. Moreover, we observe that the dataset size has no significant influence on the PG generalization ability.


We now consider fine tuning for the PG.
A first PG dataset with 80 000 positions is used to train the network. 
Then, a second PG generalization dataset, representing PG from another factory, is generated.
This dataset has a limited number of positions, $1000$, chosen randomly in the scene. This allows to mimic a random measurement process in a factory. 
The pre-trained network is then trained on this light generalization dataset. 

\begin{figure}[t]
    \centering
    \includegraphics[scale=.8]{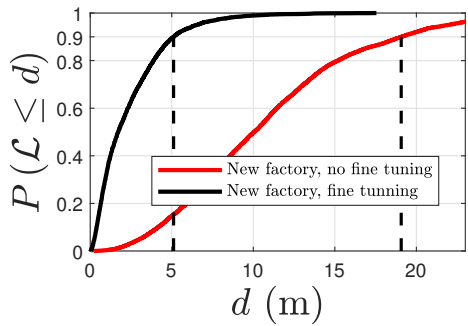}
    \caption{Transfer learning for the PG dataset in the $60\%$ clutter density scenario. The $Q(0.9)$ values are  5.1 and 19.1.}
    \label{fig:transfer_learning_PG}
\end{figure}

Results are provided in Figure~\ref{fig:transfer_learning_PG}. The fine tuning approach allows to obtain better results than:
\begin{itemize}
\item Using the neural network trained on the first dataset without fine tuning: The $Q\left(0.9\right)$ value goes down from $19.1$m to $5.1$m. 
Note that the red curve on the figure presents the same data as the one on Figure~\ref{fig:dataset_comp_gene} for the PG (60$\%$).
\item Training a neural network only on the second light dataset: The $Q\left(0.9\right)$ value goes down from more than $10$m (see Figure~\ref{fig:dataset_comp}) to $5.1$m. Moreover, obtaining a $Q\left(0.9\right)=5.1$m with an untrained network requires a dataset of size around 8000 (see Figure~\ref{fig:dataset_comp_gene}).
\end{itemize}

The latter aspect highlights that there is a clear advantage in using a network trained on another factory compared to an untrained network. This is an instance of transfer learning which enables to save 7000 measurement data.

One should also note that with fine tuning on PG data, we obtain better performance than the generalization performance of the CIR reported on Figure~\ref{fig:dataset_comp_gene}.



\section{Conclusions}\label{Conclusion}

In this paper, we considered the deep learning approach for indoor positioning in a NLoS environment.
We first presented and discussed the 3GPP InF statistical channel used to generate the datasets.
We then reported simulation results to assess the influence of some characteristics of the datasets on the positioning performance.
The five main takeaways are the following: 1-There is no clear performance difference between the PG and CIR datasets when tested on data from the same factory used to generate the training dataset, in the case of 18 BS. 
2-Unsurprisingly, the positioning accuracy obtained with the CIR is significantly more robust to a reduced number of BS than the PG. 3-The positioning error is approximately divided by two when the size of the training dataset is increased by a factor 4. 
4-Regarding the generalization performance, meaning that the testing data is from a different factory than the one used for the training, the CIR performs much better than the PG in the case without fine tuning. 5-If fine tuning with a limited number of data from the target factory is allowed, significant performance improvement (for the PG) is observed. It means that transfer learning works with statistical channels.

We emphasize that the light PG datasets offer competitive results compared to the CIR datasets with a high enough number of BS, and also require lighter neural network models (and therefore training time). 
This means that simpler radio signal than the CIR, which is the main radio signal considered in the 3GPP study, could be used for the positioning task. If only a limited number of BS is available, positioning based on the CIR is a good option.


\end{document}